\documentclass[12pt]{article}

\usepackage{sbc-template}
\usepackage{graphicx,url}

\usepackage[latin1]{inputenc}  
\usepackage{multirow}  
\usepackage[usenames,dvipsnames]{color}
\usepackage{float}
\usepackage{listings}  
\usepackage{color}
\usepackage{amsfonts}
\usepackage{subfigure}
\usepackage{colortbl}
\usepackage{booktabs}

\mathchardef\mhyphen="2D
     
\title{DCLfix: A Recommendation System for Repairing Architectural Violations}

\author{Ricardo Terra\inst{1,2}, Marco Túlio Valente\inst{1}, Roberto S. Bigonha\inst{1}, Krzysztof Czarnecki\inst{2}}

\address{Universidade Federal de Minas Gerais, Brazil
\nextinstitute
  University of Waterloo, Canada
  \email{\footnotesize \{terra,mtov,bigonha\}@dcc.ufmg.br, kczarnec@gsd.uwaterloo.ca}
}

\begin{document} 

\maketitle

\begin{abstract}
Architectural erosion is a recurrent problem in software evolution.
Despite this fact, the process is usually tackled in 
ad hoc ways, without adequate tool support at the architecture level. 
To address this shortcoming, this paper presents a recommendation system---called DCLfix---that provides 
refactoring guidelines for maintainers when tackling architectural erosion. In short,
DCLfix suggests refactoring recommendations for  violations detected after
an architecture conformance process using DCL, an architectural constraint \mbox{language}.
\end{abstract}

\section{Introduction}

Software architecture erosion is a recurrent problem in software evolution.
The phenomenon designates the progressive gap normally observed between two architectures: the {\em planned architecture} 
defined during the architectural design phase and the {\em concrete architecture} defined by the current 
implementation of the software system~\cite{ieeesw2010}. 
Although the causes for this architectural gap are diverse---ranging from conflicting requirements to deadline 
pressures---when the process is accumulated over years, architectural erosion can transform software architectures into unmanageable 
monoliths~\cite{sarkar09}.

Although several architecture conformance approaches have been proposed to detect architectural violations
(e.g., reflexion models, intensional views, design tests, query languages, and architecture description languages~\cite{ieeesw2010}),
there has been less research effort dedicated to the task of repairing such violations.
As~a~consequence, developers usually perform the task in ad hoc ways, without tool support at the architectural level.
We argue that the task of repairing architectural violations can no longer be addressed in ad hoc ways because
{\em architecture repair} is as important as {\em architecture checking}.
%, if not more important than.

To address this shortcoming, this paper presents $\tt DCLfix$, a recommendation system
that provides refactoring guidelines for maintainers to repairing architectural erosion. 
Specifically, it suggests refactoring recommendations for violations detected as the
result of an architecture conformance process using DCL, an architectural constraint language.

    The remainder of this paper is structured as follows. 
    Section~\ref{sec:tool} presents the design and implementation of the $\tt DCLfix$ tool, including background 
    and examples.
    Section~\ref{sec:relatedtools} discusses related tools and 
    Section~\ref{sec:finalremarks} presents final remarks.
    
%TODO: remove it
%\vspace{10pt}

\section{The \texttt{\textbf{DCLfix}} tool}
\label{sec:tool}

\begin{figure}[ht]\centering \includegraphics[width=9cm]{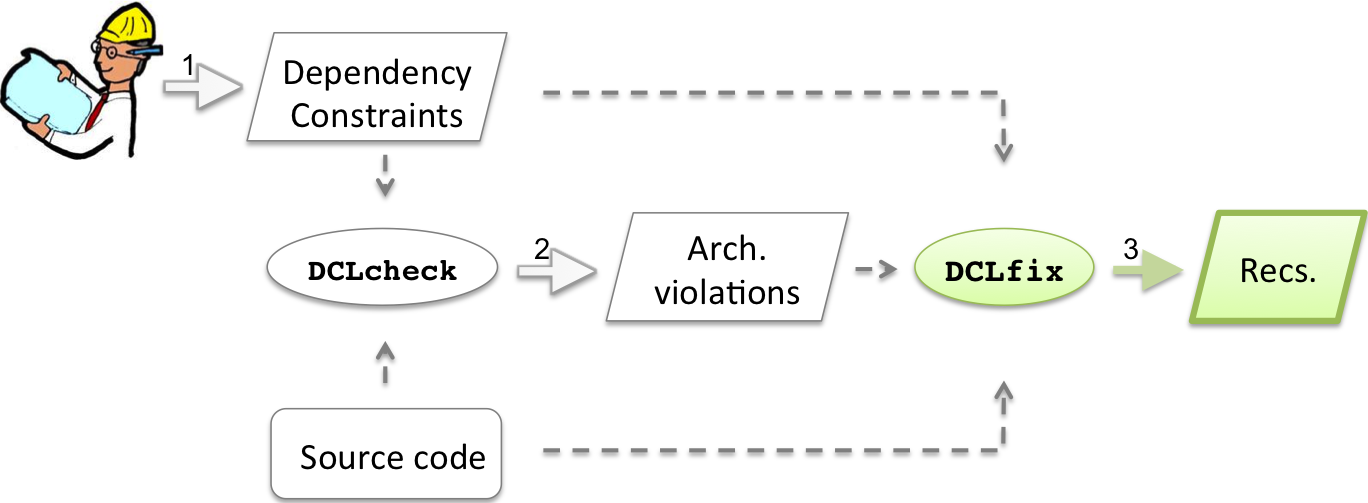} \caption{\texttt{\textbf{DCLfix}} recommendation engine}\label{fig:approachoverview} \end{figure}

As illustrated in \mbox{Figure~\ref{fig:approachoverview}}, $\tt DCLfix$---based on 
a set of DCL constraints (specified by the software architect), a set of architectural
violations (raised by the $\tt DCLcheck$ conformance tool), and the source code of
the system---provides a set of refactoring recommendations to guide the process of removing the detected violations.
For instance, in order to repair a particular violation, $\tt DCLfix$ may suggest the use of a {\em Move Class} refactoring,
including the indication of the most suitable module. 

We first provide an overview of DCL (Subsection~\ref{sec:dcl}). Next, we 
%provide an overview of 
introduce
the refactoring recommendations triggered by $\tt DCLfix$ (Subsection~\ref{sec:refrecs}).
\mbox{Finally,} we present the design and implementation of $\tt DCLfix$ 
followed by examples extracted from real case studies (Subsections \ref{sec:archinterface} and \ref{sec:examples}). 

\subsection{DCL} 
\label{sec:dcl}

The Dependency Constraint Language (DCL) is a domain-specific language that allows architects
to restrict the spectrum of structural dependencies, which can be established in object-oriented systems~\cite{dcl}.
Particularly, the language allows architects to specify that dependencies {\em only can}, {\em can only}, {\em cannot}, or {\em must} be established by specified modules. In DCL, a module is a set of classes. Moreover, it also allows architects to define the type of the dependency (e.g., $\tt access$, $\tt declare$, $\tt create$, $\tt extend$, etc.).  
In order to~explain the differences, let us assume the following constraints:

\begin{lstlisting}[emph={only,can,create,declare,depend,cannot,access,must,implement,useannotation,handle},emphstyle=\bfseries\ttfamily,
basicstyle=\small\tt,frame=no,numbers=none,emph={[2]\$java},emphstyle={[2]\tt\em},mathescape=false]
1: only Factory can-create DAO
2: Util can-depend-only JavaAPI
3: View cannot-access Model
4: DTO must-implement Serializable
\end{lstlisting}

These constraints state that only classes in the $\tt Factory$ module can create objects of classes in the $\tt DAO$ module (line 1); classes in module $\tt Util$ can establish dependencies only with classes from the Java API (line~2); classes in module $\tt View$ cannot access classes from module $\tt Model$~(line 3); and every class in the $\tt DTO$ module must implement $\tt Serializable$~(line 4).

In a previous paper~\cite{dcl}---where a complete description of DCL can be found---we have also described the $\tt DCLcheck$  tool
that checks whether DCL constraints are respected by the source code of the target system.
$\tt DCLfix$ operates on the violations detected by this tool in order to provide refactoring recommendations.

\subsection{Refactoring Recommendations} 
\label{sec:refrecs}

To provide recommendations, $\tt DCLfix$ relies on a set of 32 refactoring recommendations
formalized in previous papers~\cite{csmr2012era,icsm2012}. Table~\ref{tb:recommendation} shows 
a subset of the recommendations. 
As an example, consider a violation in which an unauthorized class~$\tt A \in M_A$ has created an object of a class~$\tt B \in M_B$.
In this case, $\tt DCLfix$ might trigger
recommendation $\tt D11$ that suggests
the replacement of the {\em new} operator with a call to the {\em get} method of a Factory~class.
%, since there is such a Factory class and $\tt A$ is allowed to access~it. 
%This recommendation addresses the fact of developers may---because unawareness or forgetfulness---create objects of classes that have a factory. 

\begin{table}[!h]
\begin{center}
\caption{Subset of Refactoring Recommendations}
\begin{scriptsize}
\begin{tabular}{p{1.4cm}p{0.15cm}p{11.2cm}r}

\bottomrule[0.8pt]\\[-0.31cm]
\multicolumn{4}{l}{\cellcolor[gray]{.95}${\tt A~cannot\mhyphen declare~B}$}\\%,\, where\, A \in M_A \wedge B \in M_B}$}\\
\toprule[0.1pt]

$\tt B \; b;\ S$ & $\Longrightarrow$ &
$
\tt
replace(\,[B], [B']\,),
\,\,\,\, \mathrm{if}\,\,\,
B' \in super(B)\,\, \wedge\,\, typecheck(\,[B'\ b\,;\, S]\,)\,\, \wedge\,\,  B' \notin M_B
$&$D1$ \\~\\[-0.25cm]

\bottomrule[0.2pt]
\multicolumn{4}{l}{\cellcolor[gray]{.95}${\tt A~cannot\mhyphen create~B}$}\\
\toprule[0.1pt]
$\tt new\ B(exp)$ & $\Longrightarrow$ &
$
\tt
replace(\,[new\ B(exp)],\, [FB.getB(exp)]\,),\,\,\, \mathrm{if}\,\,\, FB= factory(B,[exp]\,)\,\, \wedge\,\, can(A,access,FB)
$&$D11$ \\~\\

$\tt new\ B(exp)$ & $\Longrightarrow$ &
$
\tt
replace(\,[new\ B(exp)],\,[\mathrm{null}]\,),\,\,\,\, \mathrm{if}\,\,\, \overline{M_A} = \emptyset
$&$D12$ \\~\\[-0.25cm]

\bottomrule[0.2pt]
\multicolumn{4}{l}{\cellcolor[gray]{.95}${\tt A~must\mhyphen derive~B}$}\\
\toprule[0.1pt]

$\tt A$ & $\Longrightarrow$ &
$
\tt
replace (\,[A],\ [A\ derive\ B]\,),
\,\,\,\, \mathrm{if}\,\,\,
M_A = suitable\_module(A) \,\,\wedge\,\,
typecheck(\,[A\ derive\ B]\,)
$&$A3$ \\~\\

$\tt A$ & $\Longrightarrow$ &
$
\tt
move (\,A, M\,),
\,\,\,\, \mathrm{if}\,\,\,
M = suitable\_module(A)
\,\, \wedge\,\, M \not= M_A
$&$A4$ \\~\\[-0.25cm]

\bottomrule[0.2pt]
\multicolumn{4}{l}{\cellcolor[gray]{.95}${\tt A~must\mhyphen useannotation~B}$}\\
\toprule[0.1pt]

$\tt A $ & $\Longrightarrow$ &
$
\tt
replace (\,[A],\, [\,@B\;\; A\,]),
\,\,\,\, \mathrm{if}\,\,\,
M_{A} = suitable\_module(A) \,\, \wedge\,\,
target(B) = \mathrm{type} 
$&$A6$ \\~\\[-0.3cm]

      \bottomrule[1pt]
\end{tabular}
\label{tb:recommendation}
\end{scriptsize}
\end{center}
\end{table}

%Due to space restrictions, this paper does not provide the description of refactoring and 
%auxiliary functions, such as $\tt replace$, $\tt typecheck$, and $\tt suitable\_module$,
%used in Table~\ref{tb:recommendation}.
%A complete description of these functions and also the entire set of refactoring
%recommendations can be found at~\cite{icsm2012}. 

Many recommendations (e.g., $\tt A3$, $\tt A4$, and $\tt A6$) rely on the $\tt suitable\_module$ function.
This function considers a class or a module as the set of  
dependencies that it establishes with other program elements.
Based on the {\em Jaccard Similarity Coefficient}---a statistical measure for the similarity between two sets---%
it returns the module of the system with the highest similarity.
For example, a class that relies extensively on GUI types is likely to have $\tt View$
as its suitable module.
Due to space restrictions, this paper does not provide the description of other refactoring and 
auxiliary functions used in Table~\ref{tb:recommendation}, such as $\tt move$, $\tt replace$, and $\tt typecheck$.
A complete description of these functions and also the entire set of refactoring
recommendations can be found at~\cite{icsm2012}.

\subsection{Internal Architecture and Interface}
\label{sec:archinterface}

We have implemented $\tt DCLfix$ as an extension of the $\tt DCLcheck$ Eclipse plug-in.
As~illustrated in Figure~\ref{fig:dclfix}a,
$\tt DCLfix$ exploits preexisting data structures, such as the graph of existing
dependencies, the defined architectural constraints, and the detected violations.
Moreover, $\tt DCLfix$ also reuses functions implemented in $\tt DCLcheck$,~e.g.,
to check whether a type can establish a particular dependency
with another type. % (function $\tt can$).

\noindent
{The current $\tt DCLfix$ implementation %has six classes and 1,170 LOC and 
follows an 
architecture with \textcolor{black}{three} main modules:}

\begin{itemize}

\item {\em Recommendation Engine}: 
This module is responsible for determining the appropriate refactoring recommendation
for a particular violation.
More specifically, $\tt DCLfix$ has been designed as a marker resolution because
$\tt DCLcheck$ marks architectural violations on the source code (see Figure~\ref{fig:dclfix}b). 
In short, it first obtains information about the architectural violation (e.g., violated constraint
and code location),
and then, using the auxiliary functions, searches for potential refactoring recommendations.\\[-0.315cm]

\item {\em Auxiliary Functions}: This module implements the auxiliary functions used in the preconditions of
refactoring recommendations, such as checking whether the refactored code type checks ($\tt typecheck$), 
searching for design patterns (e.g., $\tt factory$), and calculating the most suitable module ($\tt suitable\_module$).\\[-0.315cm]

\begin{figure}[H]
    \centering
\begin{minipage}[b]{7cm} 
      \centering 
      \subfigure[Architecture] %\texttt{\textbf{DCLfix}}
      { 
      \raisebox{ 0mm }{ %
        \resizebox{5.3cm}{!}{\includegraphics{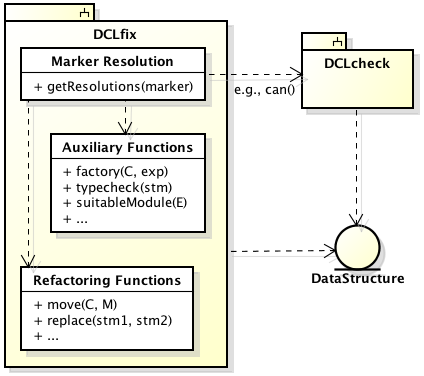}\label{fig:dclfixarchitecture}} } 
      } 
     \end{minipage}% 
    \hspace{3pt}
    \subfigure[Interface] %\texttt{\textbf{DCLfix}} 
    {
        \includegraphics[height=4.79cm]{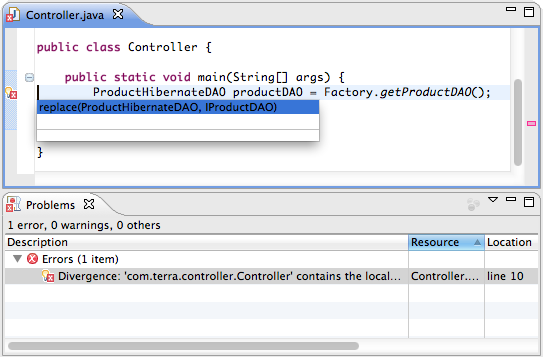} 
        \label{fig:dclfixinterface}
    }
    \caption{\texttt{\textbf{DCLfix}} architecture (2a) and interface (2b)}
    \label{fig:dclfix}
\end{figure}
%TODO: remove
%~\\[-1.1cm]

\item {\em Refactoring Functions}: This module is responsible for applying the refactorings in the source code,
e.g., the $\tt replace$ and $\tt move$ functions.
This module is still under development.
%Nevertheless, in the current stage of its implementation, this module has not been fully implemented.
\end{itemize}

As an example, consider a constraint of the form $\tt Controller\ cannot{-}depend$ $\tt HibernateDAO$. 
This constraint prevents the Controller layer from manipulating directly Hibernate {Data Access Objects} (DAOs).
Assume also a class $\tt Controller$ that declares a variable of a type $\tt ProductHibernateDAO$. 
When the developer requests a recommended fix for such violation, $\tt DCLfix$ indicates the most appropriate 
refactoring~(see Figure~\ref{fig:dclfix}b).  
The provided recommendation suggests 
replacing the declaration of the unauthorized type $\tt ProductHibernateDAO$ with its interface $\tt IProductDAO$
(which corresponds to recommendation $\tt D1$ in Table~\ref{tb:recommendation}).
This recommendation is particularly useful to handle violations due to references to a 
concrete implementation of a service, instead of its general interface.

%\subsection{Real Examples}
\subsection{Applications}
\label{sec:examples}

In our previous paper~\cite{icsm2012},
we have evaluated the application of our tool in two industrial-strength systems:
\mbox{(i)~Geplanes,} an open-source strategic management system, in which $\tt DCLfix$ triggered correct
refactoring for 31 out of 41 violations; 
(ii) TCom, a large customer care system used by a telecommunication company, 
in which $\tt DCLfix$ triggered correct refactoring for 624 out of 787 violations.

Table~\ref{tb:results} summarizes some results obtained from the evaluation of the aforementioned systems,
including the constraint description, the number of raised violations, and the triggered refactoring recommendations.
In order to illustrate the recommendations provided by $\tt DCLfix$,
we have chosen one constraint from Geplanes ($\tt GP4$) and three constraints from TCom ($\tt TC1$, $\tt TC5$, and $\tt TC9$).

\begin{table}[!ht]
\begin{center}
\begin{scriptsize}
\caption{Geplanes and TCom results}
\begin{tabular}{p{0.35cm}p{9cm}cc}
      \toprule[0.8pt]\\[-0.4cm]
      \multicolumn{2}{l}{\bf Constraint} & {\bf \# Violations}  & \multicolumn{1}{c}{\bf Correct Recs.} \\
       \midrule
	{$ \tt GP4$} & {\scriptsize Entities {\bf must-useannotation} linkcom.neo.bean.annotation.DescriptionProperty} 
	& $18$ & $\tt A6~({\displaystyle18}$ cases)\\[0.05cm]
	{$\tt TC1$} &  {\scriptsize DTO {\bf must-implement} java.io.Serializable} & $~63$ & $\tt A3~({\displaystyle50}$ cases)\\[0.05cm]   
	{$\tt TC5$} &  {\scriptsize {\bf only} tcom.server.persistence.dao.BaseJPADAO {\bf can-create} DAO} & $~13$ & $\tt D11~({\displaystyle13}$ cases)\\[0.05cm]
	{$\tt TC9$} &  {\scriptsize {\em \$system} {\bf cannot-create} Controller, DataSource} & $~~3$ & $\tt D12~({\displaystyle3}$ cases)\\[0.05cm]			
      \bottomrule[0.8pt]
\end{tabular}
\label{tb:results}
\end{scriptsize}
\end{center}
\end{table}

As a first example, constraint $\tt GP4$ states that every class in the $\tt Entities$ module must be annotated
by $\tt DescriptionProperty$. 
This constraint ensures a rule prescribed by the underlying framework in which every persistent class has to
set a description property. 
As the result of an architectural conformance process, $\tt DCLcheck$ has detected some classes in the 
$\tt Entities$ module without such annotation. 
Because these classes were located in their correct module (as certified by $\tt suitable\_module$ function),
$\tt DCLfix$ has correctly suggested adding the class-type annotation to them (rec.~$\tt A6$).

As a second example, constraint $\tt TC1$ prescribes the serialization of Data Transfer Object (DTOs).
For 50 out of 63 violations, $\tt DCLfix$ has suggested
adding the implementation of $\tt Serializable$~(rec.~$\tt A3$). 
For the other violations, $\tt DCLfix$ improperly
triggered the recommendation $\tt A4$ because 
the $\tt suitable\_module$ function considered them as $\tt Constant$
instead of $\tt DTO$ classes.
The reason is that both $\tt DTO$ and $\tt Constant$ classes rely heavily on Java's built-in types
and therefore are structurally very similar.
%(scored as incorrect by the software architect)

As another example, constraint $\tt TC5$ specifies a factory class for DAOs.
$\tt DCLcheck$ has indicated instantiations of DAO objects outside the factory. In this case, 
$\tt DCLfix$ was able to find the factory and suggested replacing the instantiation 
with a call to the factory~(rec.~$\tt D11$). 
As a last example, constraint $\tt TC9$ forbids any class of the system to create objects of $\tt Controller$ or $\tt DataSource$ classes.
In fact, these objects must be created by dependency injection techniques and thus 
no class of the system is allowed to create them.
As a result, $\tt DCLfix$ has correctly suggested the removal of the instantiation statements~(rec.~$\tt D12$).

\section{Related Tools}
\label{sec:relatedtools}

Recommendation Systems for Software Engineering (RSSEs) are ready to become part of industrial software 
developers' toolboxes~\cite{rsse2010}. Such systems usually help developers to find information and make decisions whenever they lack
experience or cannot handle all available data. 
For example, since frameworks are usually large and difficult to understand,
Strathcona~\cite{contextmatching} is a tool that recommends relevant source code fragments to help developers to use frameworks and APIs.
Our approach is also realized as a \mbox{recommendation} system, but our focus is following the planned architecture, instead of using a framework.

As another example, SemDiff~\cite{changesframework} recommends replacement methods for adapting code to a new library version,
i.e., it finds suitable replacements for framework elements that were accessed by a client program but removed as part of the framework's 
evolution. 	Analogously, $\tt DCLfix$ provides suitable replacements for implementation decisions that
denote violations in the software evolution. 

As a last example, eRose~\cite{guidechanges} identifies program elements (classes, methods, and fields)
that usually are changed together. For instance, when developers want to add a new preference to the Eclipse IDE
and then change $\tt fKeys[]$ and $\tt initDefaults()$, eRose would recommend changing also the 
$\tt Plugin.properties$ file, because, according to versioning system, they are always changed together.
However, despite of a trend towards the use of recommendation systems in software engineering, 
we are not aware of recommendation systems whose precise goal is to help developers and maintainers in tackling the architectural erosion process.

\section{Final Remarks}
\label{sec:finalremarks}

Architectural erosion is a recurrent problem in software evolution.
Although many approaches and commercial tools have been 
proposed to detect architectural violations, there has been less research effort
dedicated to the task of repairing violations.
Developers usually perform the task of fixing violations in ad hoc ways,
without tool support at the architectural level. 

To overcome these difficulties, we have developed $\tt DCLfix$---a solution based 
on recommendation system principles---that provides refactoring guidelines
for developers when repairing architectural violations. It
prevents developers to waste a long time on determining the proper fix or to introduce
new violations while fixing one. Even though the good results obtained in 
our previous evaluation with two 
industrial-strength systems~\cite{icsm2012}, we are still evaluating
$\tt DCLfix$ with other systems and our plan is to allow developers
to extend $\tt DCLfix$ with their own domain-specific refactorings recommendations.

The $\tt DCLfix$ tool---including its source code---is publicly
available at $\tt http{:}//github.com/rterrabh/DCL$.\\%
%TODO: remove
%[-0.2cm]

\noindent{\bf Acknowledgments}:~Our research has been supported by CAPES, FA\-PE\-MIG, and CNPq. %(BEX~1005/11-1).

%\scriptsize
\small
\bibliographystyle{abbrv}
\bibliography{thesisbibfile}

\end{document}